\documentstyle[10pt,epsf]{elsart}

\begin{document}

\begin{frontmatter}
\title{Low energy n+t scattering and the NN forces}
\author{F. Ciesielski, J. Carbonell and C. Gignoux}
\address{Institut des Sciences Nucl\'{e}aires
        53, Av. des Martyrs, 38026 Grenoble, France}

\begin{abstract}
The Faddeev-Yakubovsky equations have been solved in configuration space for the four nucleon system.
The n+t elastic cross section has been calculated using several realistic interactions and
a phenomenological three-nucleon force.
Special attention is devoted to the description of the observed resonant structure at $T_{lab}\approx 4$ MeV.
In the present state of the calculations, the realistic NN interactions are unable
to describe the energy region considered, i.e. n+t S- and P- waves.
Whereas the inclusion of a three-nucleon force improves the zero energy region,
the disagreement in the resonant peak remains.
This failure contrasts with the results of a pure phenomenological model.
The origin of this disagreement as well as the possible issues are discussed. 
\end{abstract}
\end{frontmatter}

%%%%%%%%%%%%%%%%%%%%%%%%%%%%%% 72 OK %%%%%%%%%%%%%%%%%%%%%%%%%%%%%%%%%% %
\section{Introduction}

The four nucleon continuum spectrum constitutes the present challenge of the few-body nuclear problem.
Its interest lies not only in the natural progression that it represents
towards the description of an increasing complexity but also in the richness of the A=4 nuclear chart itself. 
The problem implies, we believe, a qualitative jump with respect to the A=3 case.
Indeed the failure of the usual NN interactions in describing the 3N system and the 4N bound state can be 
to a large extent overcome by the inclusion of three nucleon interactions (TNI) \cite{LA_90,LA_91,C_91,GK_93.2,VKR_95}.
The success of this approach is mainly due to the strong correlation existing between
the low energy scattering observables and the three nucleon binding energy \cite{LA_90,LA_91,P_77},
which in its turn is strongly correlated to the four nucleon one \cite{TJON_75,AF_89.1,KG_92.b}.
This means that the zero and low energy three and four nucleon problem depend on a few number
of parameters which can be found in any existing TNI. 
By adjusting these parameters in order to reproduce the triton binding energy
a global satisfactory description of the 3N system and the 4N bound state is reached \cite{GK_93.2,GWHKG_96,NHKG_97}. 
The strength of the TNI depends on the underlying NN forces and
their main effect is to rescale the 3N and 4N bound states energies. Their influence
on the 3N continuum manifests itself only at relatively high energies \cite{WGHGK_98}.

However, unlike the 3N case, the 4N continuum spectrum exhibits a rich variety of resonances and thresholds
sufficiently far from the zero energy region that cannot be determined by the low energy properties.
The simple comparison between the smooth behaviour observed in the n+d elastic cross section
and the non trivial structure manifested in the n+t case illustrates this point well.
It is therefore not clear that the approach followed until now, based on a good description
of the deuteron and triton binding energies, could be successfully extended. 
A resonant state spreads over an energy region given by its width
and fully displays there the internal dynamics of the 4N system. 
Its theoretical description provides a severe test in our understanding of the nuclear forces.

Among the different scattering channels the n+t elastic case
is the simplest one and so is the obliged first step before a more ambitious program can be undertaken. 
On the one hand, it is is a pure T=1 channel free from
the additional difficulties arising when dealing with Coulomb forces.
On the other hand high quality measurement of the total n+t cross sections
are available \cite{PBS_80}.
They show a resonant structure interpreted
by the R-matrix analysis of Hale and collaborators 
as being generated by a family  of resonances \cite{TWH_92}.
The aim of this work is to investigate whether the usual realistic NN interactions
are able to account for these experimental facts.

The results we will present in what follows have been obtained  by solving
the Faddeev-Yakubovsky (FY) equations in configuration space.
The formal apparatus and details on the numerical methods
can be found in \cite{Fred_97,CC_98}.
We will restrict ourselves in this letter
to a short presentation of the formalism in the case of 
four identical particles (next section) followed
by the results and a conclusion.

%%%%%%%%%%%%%%%%%%%%%%%%%%%%%% 72 OK %%%%%%%%%%%%%%%%%%%%%%%%%%%%%%%%%%
\section{Formalism}

In the case of four identical fermions interacting via a pair-wise potential V,
the FY equations result in a set of two integrodifferential equations,
coupling two FY amplitudes K and H:
\begin{eqnarray}
(E-H_0-V)K &=&V\left[(P_{23}+P_{13})\;(-1+P_{34})\;K -
(P_{23}+P_{13})\; H\right]\cr
(E-H_0-V)H &=&V\left[(P_{13}P_{24}+P_{14}P_{23}) \;K +
P_{13}P_{24}\; H\right]  \label{FYE}
\end{eqnarray},
in which  $P_{ij}$ are the permutation operators. The wavefunction is given by:
\begin{eqnarray}
     \Psi  &=&  \Psi_{1+3} + \Psi_{2+2} \cr  
\Psi_{1+3} &=&  \left[ 1+ \varepsilon(P_{13}+P_{23}       ) \right]\;
                \left[ 1+ \varepsilon(P_{14}+P_{24}+P_{34}) \right]K\label{psi13}\\
\Psi_{2+2} &=&
\left[ 1+\varepsilon( P_{13}+ P_{23}+ P_{14}+ P_{24} ) + P_{13}P_{24}\right]\;H  \label{psi22}
\end{eqnarray}
Each FY amplitude F=K,H is considered as a function of its own set of Jacobi coordinates
$\vec{x},\vec{y},\vec{z}$ defined respectively by
\[ \begin{array}{ccc}
\vec{x}_K&=& \vec{r}_2-\vec{r}_1 \cr
\vec{y}_K&=& \sqrt{4 \over3}\left(\vec{r}_3-
{\vec{r}_1+\vec{r}_2\over2}\right)\cr
\vec{z}_K&=& \sqrt{3 \over2}\left(\vec{r}_4-
{\vec{r}_1+\vec{r}_2+\vec{r}_3\over3}\right) 
\end{array}\qquad
\begin{array}{ccc}
\vec{x}_H&=& \vec{r}_2-\vec{r}_1 \cr
\vec{y}_H&=& \vec{r}_4-\vec{r}_3 \cr
\vec{z}_H&=& \sqrt{2}\left({\vec{r}_3+\vec{r}_4\over2}-
{\vec{r}_1+\vec{r}_2\over2}\right)
\end{array}\]
and expanded in angular variables for each coordinate according to
\begin{equation}\label{KPW}
\langle\vec{x}\vec{y}\vec{z}|F\rangle=
\sum_{\alpha} \; {F_{\alpha}(xyz)\over xyz} \;Y_{\alpha} (\hat{x},\hat{y},\hat{z})  .
\end{equation}
The quantities $F_{\alpha}$ are called regularized FY components and $Y_{\alpha}$ are 
tripolar harmonics containing spin, isospin and angular momentum variables.
The label $\alpha$ holds for the set of quantum numbers defined in a coupling scheme
and includes the type of amplitude K or H.
We have chosen the j-j coupling represented by
\begin{eqnarray*}
K&\equiv&\left\{ \left[ (t_1 t_2)_{\tau_x} t_3 \right]_{T_3} t_4 \right\}_T
\otimes
\left\{ \left[ \left( l_x (s_1 s_2)_{\sigma_x} \right)_{j_x} (l_y s_3)_{j_y}
\right]_{J_3} (l_z s_4)_{j_z} \right\}_J\cr
H&\equiv&\left[ (t_1 t_2)_{\tau_x} (t_3 t_4)_{\tau_y} \right]_T \otimes
\left\{ \left[ \left( l_x (s_1 s_2)_{\sigma_x} \right)_{j_x} 
\left( l_y (s_3 s_4)_{\sigma_y} \right)_{j_y} \right]_{j_{xy}} l_z \right\}_J
\end{eqnarray*}
where $s_i$ and $t_i$ are the spin and the isospin
of the individual particles and
$T,J$ the isospin and total angular momentum of the four-body state.
Each of the $N_c=N_K+N_H$ components in the expansion (\ref{KPW})
is labeled by 12 quantum
numbers, restricted by the antisymmetry properties to the conditions
$(-)^{\sigma_x+\tau_x+l_x}=\varepsilon$ for K and
$(-)^{\sigma_x+\tau_x+l_x}=(-)^{\sigma_y+\tau_y+l_y}=\varepsilon$ for H.  

The boundary conditions for a 1+3 scattering problem are implemented by imposing
at large enough value of $z$ the Dirichlet-type condition  
\begin{eqnarray*}
K(x,y,z) &=&  t(x,y) \\
H(x,y,z) &=&  0
\end{eqnarray*}
$t(x,y)$ being the triton Faddeev amplitudes with quantum numbers
\(\left[\left(l_x(s_1 s_2)_{\sigma_x}\right)_{j_x}(l_y s_3)_{j_y}\right]_{J_3}\).
They ensure a solution which, e.g. for a relative n+t S-wave, behaves asymptotically like
\begin{eqnarray*}
K(x,y,z) &\sim&  t(x,y)\sin{(qz+\delta)}
\end{eqnarray*}
where $\delta$ is the n+t phaseshift and $q$, the conjugate momentum of the $z$-Jacobi coordinate in K-amplitudes,
is related to the center of mass n+t kinetic energy $T_{\mathrm{cm}}$ and
the physical momentum $k$ by
\(T_{\mathrm{cm}}={3\over4}T_{\mathrm{lab}}={\hbar^2\over m}q^2={2\over3}{\hbar^2\over m}k^2\)

%%%%%%%%%%%%%%%%%%%%%%%%%%%%%%%%%%%%%%%%%%%%%%%%%%%%%%%%%%%%%%%%%%%%%%%%%%%%
\section{Results}

The results we will present concern the $^4$He bound state and the n+t elastic scattering in the relative S- and P- waves. 
They have been obtained with the AV14 \cite{AV14_84}, NIJM-II and REID-93 \cite{NIJ_93} NN interactions.
They turned out to provide very similar predictions in the observables we have considered and
we have chosen to give in more detail only the results of AV14.
They will also be compared to the predictions of the S-wave phenomenological
MT I-III model published in \cite{CC_98}.

For $^4$He without Coulomb forces
we obtain the binding energies (B) and r.m.s radius ($\bar{r}$) displayed in Table~\ref{Tab_BS}.
The NN interaction is limited to $^1S_0$, $^3S_1$, $^3D_1$ partial waves 
and the number of FY components in expansion (\ref{KPW}) 
is limited to $N_c=5+5$ and $N_c=15+9$, the same as those listed in \cite{KG_92.b}.
The results provided by the different potentials are relatively close to each other but still far from
the Coulomb corrected experimental value $B=29.0$ MeV.
For AV14 they are in good agreement with the momentum space calculations,
e.g. $B^{5+5}=23.36$ MeV and $B^{15+9}=23.77$ MeV from \cite{KG_92.b},
while more complete calculations give as converged values 
$24.2$ MeV \cite{PPCPW_97,V_98} and $24.7$ MeV  \cite{GKWHGM_95.3}. 
For the two remaining potentials we present here the first calculations.
%%%%%%%%%%%%%%%%%%%%%%%%%%%%%%%%
\begin{table}[htb]
\caption{$^4$He binding energy B (MeV) and r.m.s. radius $\bar{r}$ (fm).}\label{Tab_BS}
\begin{tabular}{ccccc}\hline
& \multicolumn{2}{c}{$N_c=5+5$} & \multicolumn{2}{c}{$N_c=15+9$} \\ 
        & B     & $\bar{r}$    & B     & $\bar{r}$  \\ \hline
AV14    & 23.34 & 1.56 & 23.81 & 1.54 \\
NIJM~II & 23.39 & 1.54 & 23.86 & 1.53 \\
REID~93 & 23.65 & 1.53 & 24.12 & 1.52 \\\hline
\end{tabular}
\end{table}
%%%%%%%%%%%%%%%%%%%%%%%%%%%%%%

In the scattering calculations we will expand the different $J^{\pi}$ n+t states
in terms of the  asymptotic hamiltonian eigenstates $|L,S;J^{\pi}>$
in which $L$ is the n+t relative angular momentum and $\vec{S}=\vec{s}_n+\vec{s}_t$ the total spin.
The following $J^{\pi}$ states will be considered:
\begin{eqnarray}
|0^{+}\rangle=&       &|0,0;0^{+}\rangle                 \cr
|1^{+}\rangle=&c_{1^+}&|0,1;1^{+}\rangle + d_{1^+}|2,1;1^{+}\rangle \cr
|0^{-}\rangle=&       &|1,0;0^{-}\rangle                  \cr
|1^{-}\rangle=&c_{1^-}&|1,1;1^{-}\rangle + d_{1^-}|1,0;1^{-}\rangle \cr
|2^{-}\rangle=&c_{2^-}&|1,1;2^{-}\rangle + d_{2^-}|3,1;2^{-}\rangle     \label{ntstates}
\end{eqnarray}
where the coefficients $c$ and $d$, fixed by the dynamics, make them eigenstates of the S-matrix.
These states are the only relevant ones for the low scattering energies considered.
Moreover the coupling $L\leftrightarrow L+2$ between the different asymptotics in the $J^{\pi}=1^{+},2^{-}$
states turned to be very small and has been neglected in the calculations presented below,
which are thus restricted to the lowest angular momentum component in the asymptotic channel.
For the remaining $J^{\pi}=1^{-}$ coupled state we present the two corresponding eigenphase-shifts.

%%%%%%%%%%%%%%%%%%%
\subsection{S-waves}

The  phase-shifts for the first positive parity states ($J^{\pi}=0^+,1^+$)
with the AV14 interaction are given in the first two columns of Table \ref{Tab_depha_SP}.
They have been obtained with the NN potential acting only upon the $^1S_0$, $^3S_1$, $^3D_1$ partial waves
and limiting the expansion (\ref{KPW}) to $l_y,l_z\leq 2$ except in the K amplitudes where $l_z\leq 1$. 
That makes for $J^{\pi}=0^+(1^+)$ a number of FY components equal to $N_c=19(48)$.
This criterion corresponds to the $N_c=15+9$ calculations for the $^4$He 
(see Table \ref{Tab_BS}) and $N_c=5$ Faddeev amplitudes for the triton asymptotic state.
This triton has a binding energy $B_3=7.43$ MeV, to be compared with
the fully converged value $B_3=7.67$ MeV obtained with $N_c=34$ \cite{CPFG_85}.

\begin{table}[hbt]
\caption{AV14 n+t S- and P-wave phase-shifts (degrees) as a function
of center of mass kinetic energy $T_{\mathrm cm}$ (MeV).}\label{Tab_depha_SP}
\begin{tabular}{ccccccccc}
\hline
$T_{\mathrm cm}$& 0$^+$& 1$^+$ &0$^+$ (TNI)& 1$^+$ (TNI)& 0$^-$ & 1$^-$ & 1$^-$ & 2$^-$  \\\hline
0.05         & 169.5   & 170.8 &  170.3    & 171.4      &  0.09 & 0.15  & 0.19  & 0.11 \\
0.5          & 147.9   & 151.7 &  150.1    & 153.4      &  2.2  & 4.3	& 5.9   & 4.0  \\
1.0          & 135.8   & 140.8 &  138.7    & 143.1      &  5.9  & 10.5  & 16.1  & 10.9 \\
2.0          & 120.5   & 126.9 &  124.1    & 129.7      & 13.8  & 20.6  & 35.9  & 25.0\\
3.0          & 110.2   & 117.2 &  114.0    & 120.3      & 21.2  & 26.6  & 47.8  & 35.4\\
4.0          & 102.3   & 109.8 &  106.3    & 113.0      & 27.4  & 30.1  & 53.5  & 41.9 \\
5.0          &  95.5   & 103.7 &  99.9     & 106.9      & 32.3  & 32.1  & 56.3  & 45.8\\
6.0          &   -     & -     &  94.3     & 101.6      & 34.8  & 35.1  & 59.6  & 48.7\\
\hline
\end{tabular}
\end{table} 

The low energy parameters are given in Table \ref{Tab_lep} for several potentials.
The scattering lengths have been calculated directly at zero energy
whereas the effective range is extracted from fitting the results of Table \ref{Tab_depha_SP}.
One can see, as in the $^4$He bound state, very similar results for the potentials considered.
Some dependence on the maximum value of $l_z$ is observed.
The values for AV14 are in agreement with those obtained
with a variational approach $a(0^+)=4.32$ fm and $a(1^+)=3.80$ fm \cite{V_98}.

%%%%%%%%%%%%%%%%%%%%%
\begin{table}[hbt]
\caption{Singlet, triplet and coherent n+t scattering lengths $a(0^+),a(1^+),a_c$ and effective range $r_0$
 (fm). The zero energy cross section $\sigma_0$ is in fm$^2$. }\label{Tab_lep}
\begin{tabular}{ccc|cccccc}\hline
&\multicolumn{2}{c|}{$l_z=0$}&\multicolumn{6}{c}{$l_z=0,1$}\\\cline{2-9}
&$a(0^+)$&$a(1^+)$&$a(0^+)$&$r_0(0^+)$&$a(1^+)$&$r_0(1^+)$&$a_c$&$\sigma_0$\\\hline
NIJM~II &4.34&-   &4.31&-   &3.76&-    & 3.90 & 191.6 \\
AV14    &4.34&3.87&4.31&2.08&3.79&1.76 & 3.92 & 193.7 \\
AV14+TNI&-   &-   &3.99&1.95&3.53&1.71 & 3.65 & 167.5 \\\hline
\end{tabular}
\end{table} 

\begin{figure}[htb]
\begin{center}
\epsfxsize=14.0cm\mbox{\epsffile{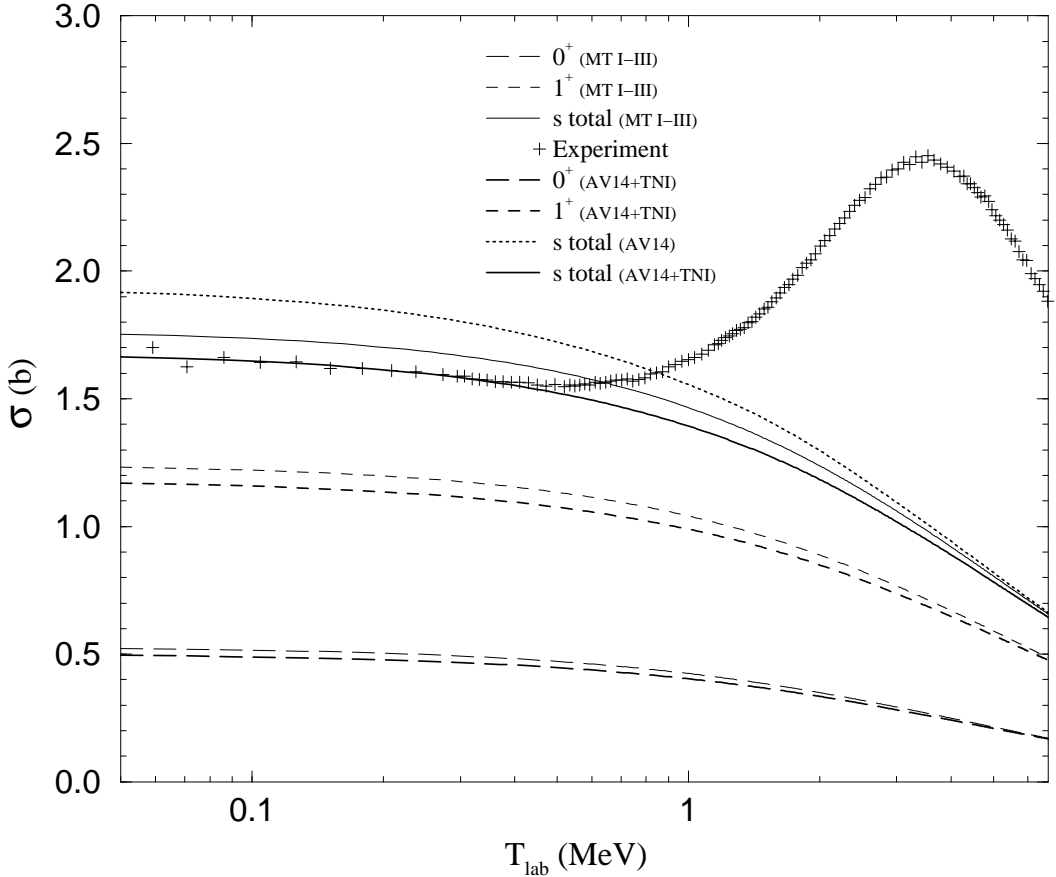}}
\caption{n+t S-wave cross section with MT I-III, AV14, and AV14+TNI interactions.}
\label{nt_s}
\end{center}
\end{figure}

The S-wave n+t elastic cross section is displayed in Figure~\ref{nt_s} 
for AV14 (dotted thick line) and compared to the experimental values taken from \cite{PBS_80}.
We have included for comparison the results obtained with MT I-III potential (solid thin line) \cite{CC_98}.
The separated singlet (long-dashed) and triplet (short-dashed) contributions are also plotted.
The disagreement with the experimental data is dramatic although expected due to the strong
correlation between the 4N binding energy (far from its experimental value) and the N+3N scattering length.
This disagreement can be overcome as in the 3N case by including a three nucleon interaction.
Several TNI models exist in the literature \cite{CS_79,CKR_83,CPW_83,SPW_86,W_91}.
In all of them, some parameters have had to be adjusted in order
to reproduce the experimental 3N, and eventually 4N, binding energy.
In view of that we have used the phenomenological model having the form
\begin{equation}\label{GRTNI}
 W(\rho)=W_r {e^{-2\mu\rho}\over\rho}-W_a {e^{-\mu\rho}\over\rho}\qquad , \qquad \rho=\sqrt{x^2+y^2} 
\end{equation}
The parameters have been chosen to reproduce the experimental triton binding energy $B_3=8.48$ MeV.
With the parameter set $W_r=500$ MeV, $W_a=174$ MeV, $\mu=2.0$ fm, one has
also a satisfactory agreement for the 4N binding energy $B_4=29.0$ MeV and
the following n+t low energy parameters: $a_0=4.0$ fm, $a_1=3.53$ fm and $\sigma(0)=167.5$ fm$^2$.
We notice that the parameter set used is far from being unique
and that the four-body calculations have been performed with $N_c=5+5$ FY components.
The corresponding n+t S-wave cross section has also been calculated
and is plotted in Figure~\ref{nt_s} (solid thick line).
The zero energy cross section is now in agreement with the experimental points.
We notice also that the values of the scattering lengths thus obtained are
close to those provided by the variational CHH method, $a_0=4.12$ fm, $a_1=3.59$ fm and $\sigma(0)=172.4$ fm$^2$,
adding to the AV14 NN potential the more sophisticated Urbana TNI-models \cite{V_98}.
This seems to prove that, at least for the observables under investigation,
the essential role of a three nucleon force is just to provide the right three ($B_3$) and four body ($B_4$)
binding energies and  consequently ensure acceptable n+t low energy parameters.

%%%%%%%%%%%%%%%%%%%
%%%%%%%%%%%%

The experimental situation concerning the n+t scattering lengths
is not very well established (see Table \ref{Tab_explep}).
Only the zero energy cross section $\sigma^{exp}(0)=1.70\pm0.03$ b \cite{PBS_80} seems to be a firm result.
The coherent scattering length $a_c$  from \cite{SBP_80} is based on 
the assumption $a_1/a_0=0.92\pm0.04$ and the deduced values for $a_0$ and $a_1$
are thus not purely experimental results.
The more recent measurement \cite{RTWW_85} gives $a_c=3.59\pm0.02$,
a value just compatible with the preceding one,
but the two solution sets inferred for the scattering lengths are
too far from any theoretical prediction to take them without some care.
The values given in \cite{HDSBP_90} are in fact extracted from p-$^3$He data
via a Coulomb corrected R-matrix analysis.
The results of our calculations together with the MT I-III predictions
tends to favour values close to $a_0=4.0-4.1$ fm, $a_1=3.53-3.60$ fm and $a_c=3.65-3.70$,
i.e. the older experimental result of \cite{SBP_80}.
It would be however of a great interest
to have a definite experimental conclusion on the $a_c$ value.

\begin{table}[hbt]
\caption{Experimental coherent ($a_c$), singlet ($a_0$) and triplet ($a_1$)
 n+t scattering lengths (fm).}\label{Tab_explep}
\begin{tabular}{cccl}
\hline
$a_c=(3a_1+a_0)/4$  &    $a_0$       &  $a_1$ & ref.	    	       \\\hline
3.68$\pm$0.05       & 3.91$\pm$0.12  & 3.60$\pm$0.1   & \cite{SBP_80}       \\ 
3.59$\pm$0.02       & 4.98$\pm$0.29  & 3.13$\pm$0.11  & \cite{RTWW_85}-I    \\ 
                    & 2.10$\pm$0.31  & 4.05$\pm$0.09  & \cite{RTWW_85}-II   \\ 
3.607$\pm$0.017     & 4.45$\pm$0.10  & 3.32$\pm$0.02  & \cite{HDSBP_90}     \\
\hline
\end{tabular}
\end{table}
 
%%%%%%%%%%%%%%%%%%%
\subsection{P-waves}

The phase-shifts for the $J^{\pi}=0^-$, $1^-$, $2^-$ states, are given in Table~\ref{Tab_depha_SP}.
The two $J^{\pi}=1^-$ columns are the eigenphase-shifts corresponding to the two
coupled asymptotics states in (\ref{ntstates}).

%%%%%%%%%%%%%%%%%%%%%%%%%%%%%%%%%%%%%%%%%%%%%%%%%%%%%%%%%%%%%%%%%%%%
\begin{figure}[hbtp]
\begin{center}\epsfxsize=14cm\mbox{\epsffile{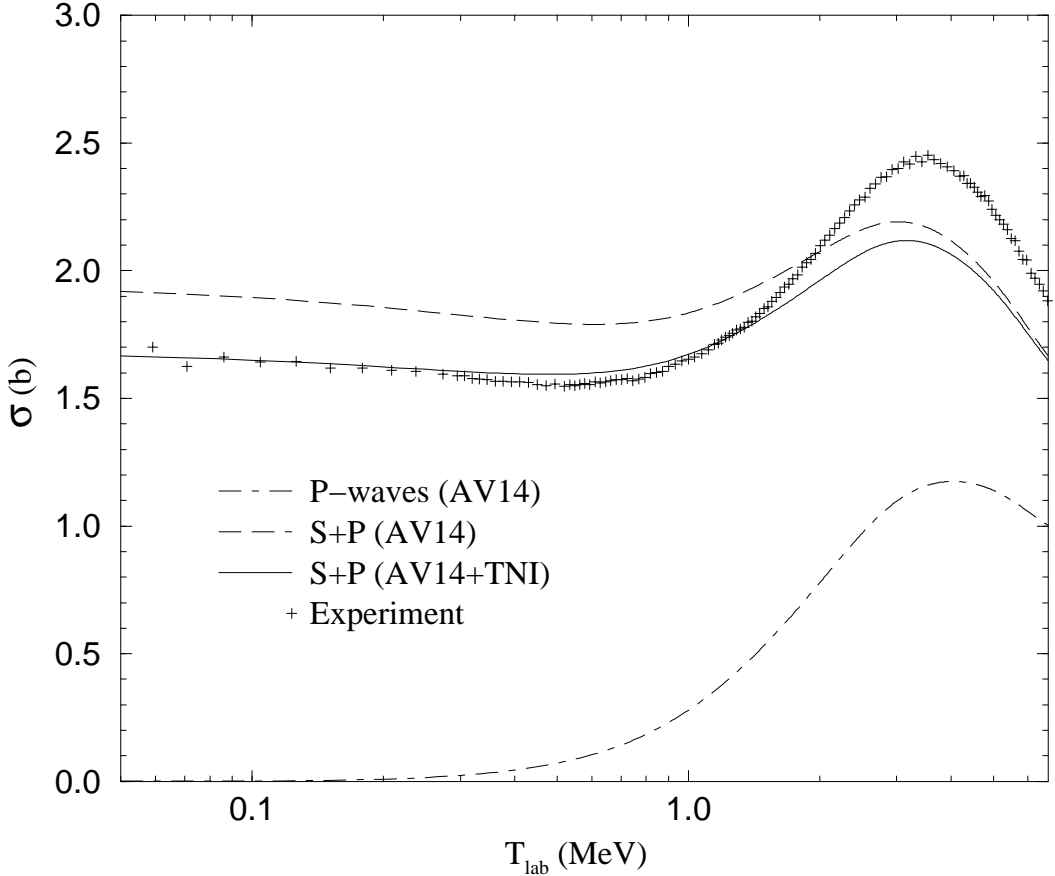}}\end{center}
\caption{n+t cross section with AV14 with and without three nucleon interaction.}\label{TOTALAV14}
\end{figure}
%%%%%%%%%%%%%%%%%%%%%%%%%%%%%%%%%%%%%%%%%%%%%%%%%%%%%%%%%%%%%%

\begin{figure}[hbtp]
\begin{center}\epsfxsize=14cm\mbox{\epsffile{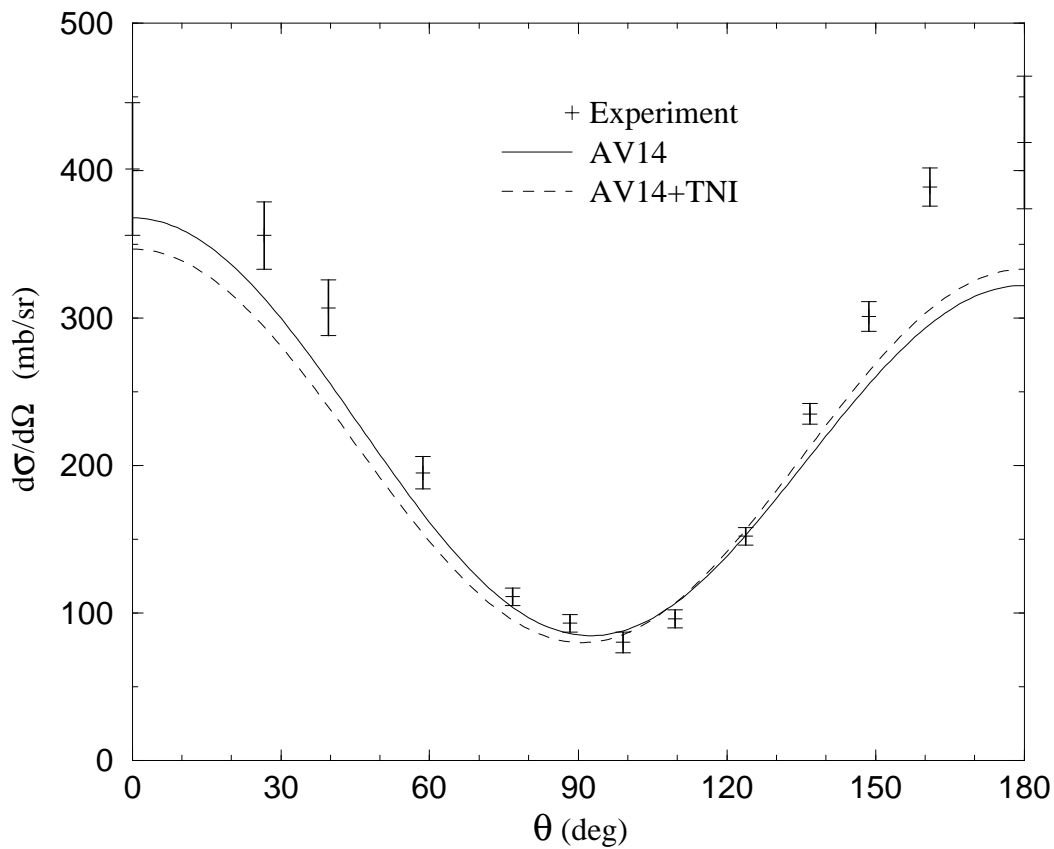}}\end{center}
\caption{The n+t differential cross section at $T_{\mathrm{lab}}=3.5$ MeV obtained with
 AV14 alone and including three-nucleon interaction are compared to experimental data.}\label{DCS}
\end{figure}

The resulting n+t elastic cross section are displayed in Figure~\ref{TOTALAV14}.
It appears that the results obtained with AV14 (dashed line) are unable to account for the experimental data. 
By removing the S-wave part in the cross section one finds that $\approx$ 1/3 of the P-wave contribution
 is missed in the calculations.  
This is in contrast with the MT~I-III results, which provides an almost perfect description of the resonant region \cite{CC_98}.
It is worth noticing that in the presence of the phenomenological three-nucleon forces (\ref{GRTNI}) this quantitative
disagreement remains and is even worse (solid line).
Whereas the TNI sensibly rescales the zero-energy cross section, 
its doesn't have any significant effect on the P-wave phase-shifts.
The differences between both cross sections, visible in Figure~\ref{TOTALAV14}, are
essentially due to the n+t S-wave contribution.  
The small influence of the three nucleon interactions in the low energy scattering observables
seems to be a general property of the existing models \cite{WGHGK_98}
rather than a consequence of the simplified TNI we have considered in (\ref{GRTNI}).
A similar disagreement can be observed in the differential cross section at neutron laboratory energy 
$T_{lab}=3.5$ MeV taken from \cite{SCS_60} and displayed in Figure~\ref{DCS}.

%%%%%%%%%%%%%%%%%%%%%%%%%%%%%%%%%%%%%%%%%%%%%%%%%%%%%%%%%%%%%%%%%
%%%%%%%%%%%%%%%%%%%%%%%%%%%%%%%%%%%%%%%%%%%%%%%%%%%%%%%%%%%%%%%%%%

\begin{figure}[hbtp]
\begin{center}\epsfxsize=14cm\mbox{\epsffile{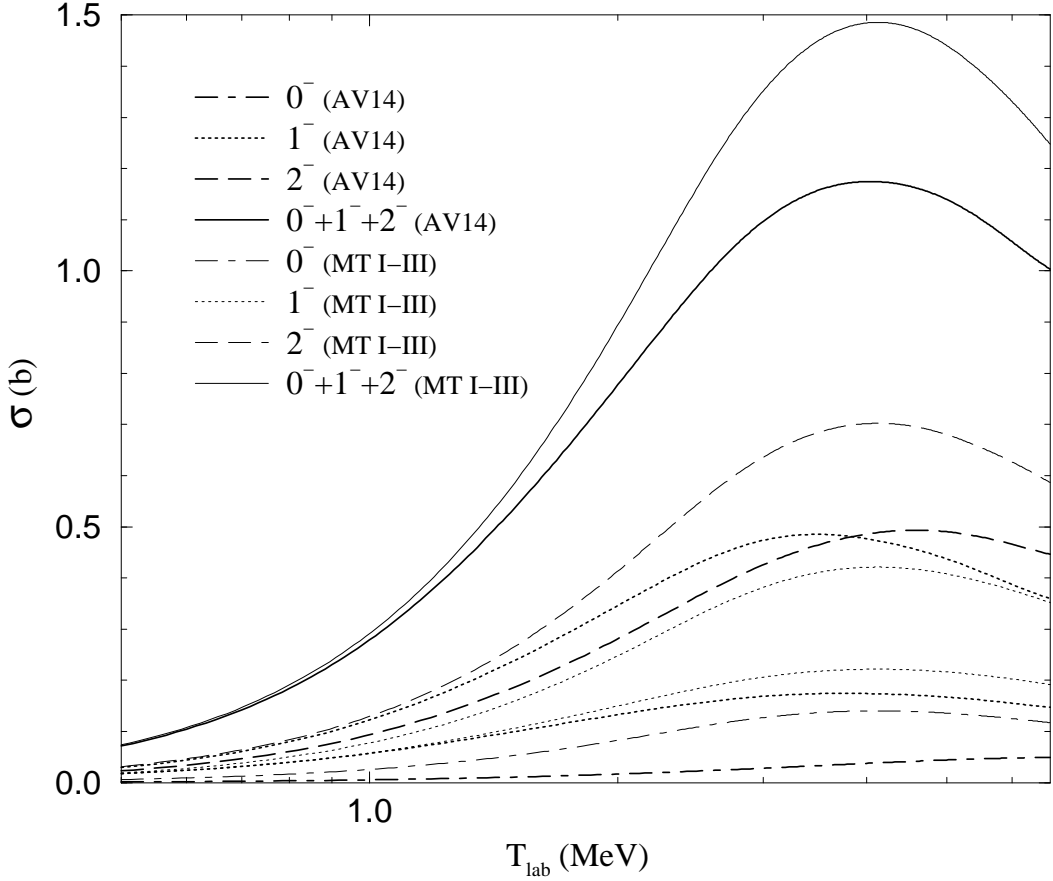}}\end{center}
\caption{Comparison of the cross sections obtained with
MT~I-III and AV14 for each partial wave.}\label{COMPMTAV}
\end{figure}

The comparison between the different P-waves contributions given by MT~I-III and AV14 potentials
is done in Figure~\ref{COMPMTAV} and shows that:\par
(i) the degeneracy between the $J^{\pi}=0^-,1^-,2^-$ states 
 existing in the MT~I-III potential no longer holds in the case of AV14
as a consequence of its tensor and $\vec{l}\cdot\vec{s}$ terms.
This splitting broadens the sum of these three contributions, which present their maximum separately
and results in a lower value for the cross section.\par
(ii) there is additionally a real lack in their separated contributions, specially in the 
$0^-$ and the statistically enhanced $2^-$ partial wave.
%%%%%%%%%%%%%%%%%%%%%%%%%%%%%%%%%%%%%%%%%%%%%%%%%%%%%%%%%%%%%%%%%

\begin{figure}[hbtp]
\begin{center}\epsfxsize=14cm\mbox{\epsffile{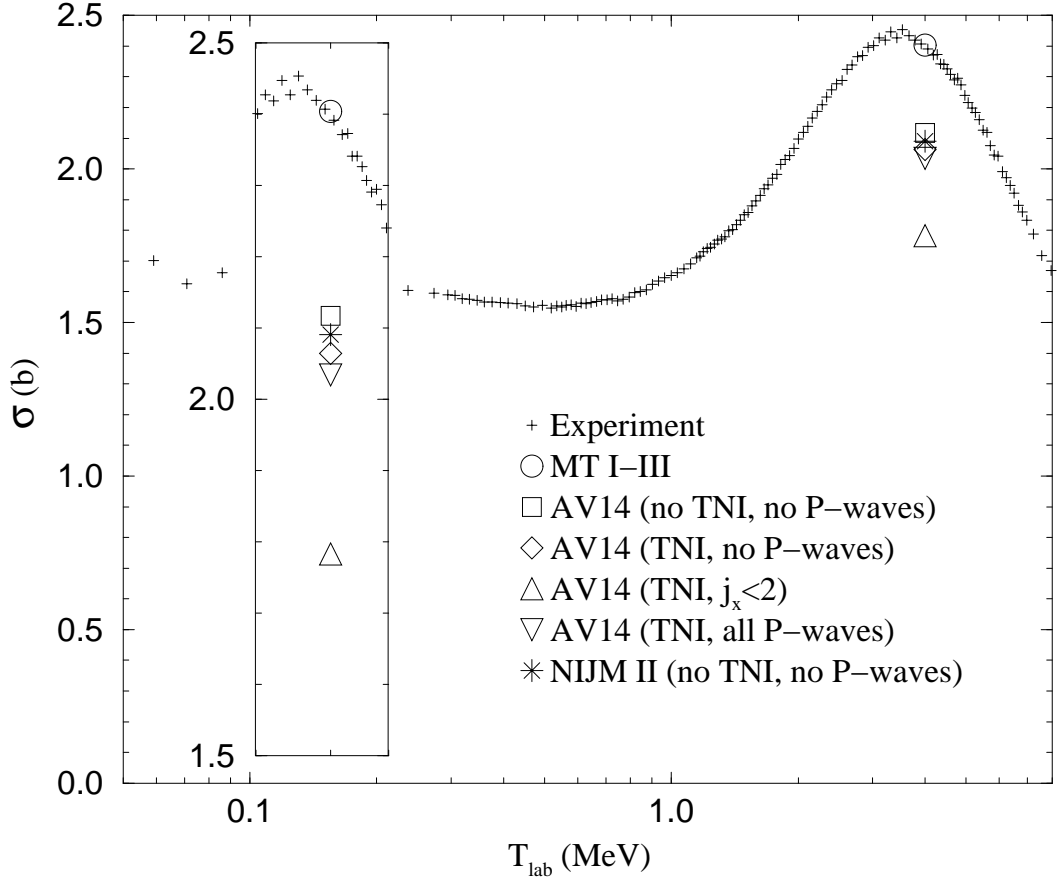}}\end{center}
\caption{Summary of different calculations
for the n+t elastic cross section at 4~MeV laboratory incident energy.}\label{EVOLSIGMA}
\end{figure}

A possible influence of P-wave NN interaction has been checked in \cite{CCG_97}.
The $^1P_1$, $^3P_0$, $^3P_1$ NN
partial waves were added to $^1S_0,^3S_1,^3D_1$, thus completing the $V^{j=1}_{NN}$ shell,
but the corrections were found to go in the wrong direction.
Since, we have included the $^3P_2-^3F_2$ components and we present in Table~\ref{DEPHACOMP}
the phase-shifts at $T_{\mathrm{cm}}=3$~MeV.
The corresponding elastic cross sections, togheter
with the MT I-III and NIJM II predictions, are given in Table~\ref{TABSIGMA} and plotted in Figure~\ref{EVOLSIGMA}.
Even if the relative weights of the different partial waves are modified,
the disagreement noticed so far still remains. 
The inclusion of the different NN P-waves interactions
has no global effect in the cross section. This conclusion 
is in agreement with the extensive study done for 3N continuum case (see e.g. Fig. 6 in \cite{GWHKG_96})
in which the $V^{j\leq1}_{NN}$ was shown to give satisfactory results at the considered low energies.

Another reason for such a disagreement, suggested by the success of the MT I-III model, could be found
in the NN S-wave tensor force itself. In particular we managed to increase the n+t P-waves
phase-shifts and consequently the total cross section by weakening the $^3S_1-^3D_1$ coupling
in benefit of the $^3S_1$ potential, keeping the same triton binding energy all along. 
The tensor force, firmly established in the deuteron and the NN observables,
seems overestimated in the more compact systems and
responsible for the underbinding problem in the 3N and 4N bound states.
A possible solution could be found by suppressing its the short range part.

%%%%%%%%%%%%%%%%%%%%%%%%%%%%%%%%%%%%%%%%%%%%%%
\begin{table}[hbt]
\caption{n+t P-waves phase-shifts (degrees) at $T_{\mathrm{cm}}$=3~MeV, 
with different truncations of the NN interaction.}\label{DEPHACOMP}
\begin{tabular}{lcccc}\hline
NN partial waves & 0$^-$ & 1$^-$ & 1$^-$ & 2$^-$ \\\hline
$^1S_0,~^3S_1,~^3D_1$
& 21.2 & 26.6 & 47.8 & 35.4 \\
$^1S_0,~^3S_1,~^3D_1$~+~$^1P_1,~^3P_0,~^3P_1$
& 24.1 & 20.0 & 37.5 & 32.7 \\
$^1S_0,~^3S_1,~^3D_1$~+~$^1P_1,~^3P_0,~^3P_1,~^3P_2,~^3F_2$
& 25.1 & 22.8 & 38.0 & 41.4 \\\hline
\end{tabular}
\end{table} 

%%%%%%%%%%%%%%%%%%%%%%%%%%%%%%%%%%%%%%%%%%%%%%%%%%%%%%%%%%%%%%%%%%%%%%%%
\begin{table}[hbt]
\caption{n+t elastic cross section $\sigma$ (barns) at $T_{\mathrm{cm}}$=3~MeV  for different potentials.}\label{TABSIGMA}
\begin{tabular}{lc}\hline
         & $\sigma$ (b) \\\hline
exp. & 2.40 \\
MT~I-III & 2.40 \\
AV14 (no TNI, no P-waves) & 2.12 \\
AV14 (TNI, no P-waves) & 2.06 \\
AV14 (TNI, $j_x<2$) & 1.78 \\
AV14 (TNI, all P-waves) & 2.03 \\
NIJM~II (no TNI, no P-waves)& 2.09 \\
\hline
\end{tabular}
\end{table} 

%%%%%%%%%%%%%%%%%%%%%%%%

Under the restrictions in which they have been obtained, i.e. limited number of amplitudes in expansion (\ref{KPW}),
the preceding results indicate that the
AV14 interaction, limited to NN S- and P-waves, cannot reproduce the n+t elastic cross section.
The calculations performed with other realistic local potentials lead to very similar results for the scattering 
lengths (Table~\ref{Tab_lep}) as well as for the  phase-shifts (Table~\ref{DEPHAAVNIJM}).
The qualitative behaviour of AV14 seems thus to be a general feature of the modern realistic NN potentials
at least in their local form.
On the other hand the recent results obtained by Fonseca \cite{CCGF_98}
reach similar conclusions in a totally independent way and extend them extensible to the non local Bonn B potential.

\begin{table}[hbt]
\caption{Comparison of the n+t P-waves phase-shifts (degrees) at $T_{\mathrm{cm}}$=3~MeV with AV14 and NIJM~II. 
The NN interaction is restricted to its $^1S_0$, $^3S_1$, $^3D_1$ components.}\label{DEPHAAVNIJM}
\begin{tabular}{ccccc}\hline
NN potential & 0$^-$ & 1$^-$ & 1$^-$ & 2$^-$ \\\hline
AV14 & 21.2 & 26.6 & 47.8 & 35.4 \\
NIJM~II & 20.6 & 26.0 & 46.0 & 34.1 \\\hline
\end{tabular}
\end{table} 

%\bibliographystyle{unsrt}

%%%%%%%%%%%%%%%%%%%%%%%%%%%%%% 72 OK %%%%%%%%%%%%%%%%%%%%%%%%%%%%%%%%%%%
\section{Conclusion}

The solutions of Faddeev-Yakubovsky equations in configuration space for the low energy 
neutron-triton scattering have been obtained by using several realistic NN interactions.

The results obtained for the n+t scattering length and elastic cross section are compared to the existing data.
The interactions considered, at least when limited
to the NN S- and P-waves, fail in describing both the zero energy cross section 
and the n+t resonant structure observed at neutron $T_{\mathrm{lab}}\approx 3.5$ MeV.
A phenomenological three-nucleon interaction has been included. 
It provides overall agreement for $^3$H and $^4$He binding energies
as well as for the n+t S-wave scattering region.
Its effect is however negligible in the resonant region and the disagreement remains.
Including NN P-wave interactions doesn't substantially modify this result.

On another hand the success obtained by using simple S-wave spin dependent potentials \cite{CC_98}
leads us to think that the combined effect of $V_{NN}+V_{NNN}+\ldots$
forces results in a simplified effective interaction, as happens in the nuclear structure calculations,
and shows how little effectives the existing OBEP models are in describing the few nucleon systems.

These results need to be confirmed and the convergence in expansion (\ref{KPW}) pushed further.
In their present state they suggest a basic inadequacy of the existing NN interactions in describing the n+t
resonant states, a problem which can hardly be corrected in terms of three nucleon forces.

%%%%%%%%%%%%%%%%%%%%%%%%%%%%%%%%%%%%%%%%%%%%%%%%%%%%%%%%%%%%%%%%%%%%%%%%%%%%%%%%%%%%%%%%%%%%%%%%%%%%%%
\ack{
The numerical calculations have been performed
with the Cray-T3E of the CGCV (CEA) and IDRIS (CNRS).
We acknowlegde the staff members of both computer centers for their constant help.}

%%%%%%%%%%%%%%%%%%%%%%%%%%%%%% 72 OK %%%%%%%%%%%%%%%%%%%%%%%%%%%%%%%%%%%

\end{document}